\documentclass[prb,twocolumn,superscriptaddress,footinbib,showpacs]{revtex4}
\usepackage{epsf}

\newcommand{\bq}{\begin{equation}}
\newcommand{\eequ}{\end{equation}}
\newcommand{\bqa}{\begin{eqnarray}}
\newcommand{\eqa}{\end{eqnarray}}
\newcommand{\nn}{\nonumber}
\newcommand{\ms}[1]{\mbox{\scriptsize #1}}

\newcommand{\la}{\langle}
\newcommand{\ra}{\rangle}

\usepackage{graphicx}
\begin{document}

\preprint{LAUR-03-1275}

\title{Feedback cooling of a nanomechanical resonator
                    \vbox to 0pt{\vss
                    \hbox to 0pt{\hskip+45pt\rm \small LAUR-03-1275\hss}
                    \vskip 25pt}}

\author{Asa Hopkins}
\affiliation{T-8, Theoretical Division, Los Alamos National
             Laboratory, Los Alamos, New Mexico 87545.} 
\affiliation{Norman Bridge Laboratory of Physics 12-33, 
             California Institute of Technology, Pasadena, CA 91125.} 
\author{Kurt Jacobs}
\author{Salman Habib}
\affiliation{T-8, Theoretical Division, Los Alamos National
             Laboratory, Los Alamos, New Mexico 87545.} 
\author{Keith Schwab}
\affiliation{Laboratory for Physical Sciences, College Park, Maryland
             20740.} 

\begin{abstract}
  Cooled, low-loss nanomechanical resonators offer the prospect of
  directly observing the quantum dynamics of mesoscopic systems.
  However, the present state of the art requires cooling down to the
  milliKelvin regime in order to observe quantum effects. Here we
  present an active feedback strategy based on continuous observation
  of the resonator position for the purpose of obtaining these low
  temperatures. In addition, we apply this to an experimentally
  realizable configuration, where the position monitoring is carried
  out by a single-electron transistor. Our estimates indicate that
  with current technology this technique is likely to bring the
  required low temperatures within reach.
\end{abstract}

%\pacs{85.85.+j,85.35.Gv,03.67.Pp,45.80.+r}
\maketitle

\section{Introduction}

Nanomechanical resonators are now being built with quality factors in
the range, $Q\approx 10^{4}$, and resonance frequencies of up to
several hundred MHz~\cite{Experiments}. The ground state energy of
these devices can correspond to temperatures in the milliKelvin range.
As a result, the observation of quantum behavior in these devices is
becoming a real possibility~\cite{Armour}.  To detect such behavior,
the resonator must be sufficiently cold; since a quantum harmonic
oscillator driven by thermal noise behaves as a classical oscillator
driven by thermal noise, one must ensure that the signatures of
quantum effects are not swamped by the thermal behavior.  The approach
taken so far to achieve low temperatures is to place the resonator in
a refrigerator. However, cooling very small devices in this way is
inherently inefficient in that the system becomes weakly coupled to
the thermal bath. Here we explore the possibility of using feedback
control to effect `active' cooling of the resonator, in order to cool
below the possible limits set by the `passive' refrigeration
technique.

To perform such feedback cooling the resonator must be monitored, and
the result fed back in real time to affect the dynamics.  A practical
method of performing a continuous measurement of the position of the
resonator is to use a Single-Electron Transistor
(SET)~\cite{Korotkov94,Hanke94,Shnirman}.  To measure the position of
the resonator one locates the central island of the SET next to the
resonator. When the resonator is charged, and the SET is biased so
that current flows through it, changes in the resonator's position
alter the potential on the central island, which in turn changes the
current. The current therefore provides a continuous measurement of
the position of the resonator, and this is just what is required for
implementing a linear feedback cooling
algorithm~\cite{Doherty99,Jacobs}. A feedback force can be applied by
applying a voltage to a gate capacitively coupled to the resonator,
and adjusting the voltage so as to damp the resonator (see Fig.~1), or
by passing a variable current through the oscillator in the presence
of a fixed external magnetic field. We will analyze the first system,
although the results should apply to the second as well.  In our
analysis we will use the theory of the dc-SET. While an experiment
would most likely use a radio-frequency SET, the characteristic
frequency of a SET is typically of the order of $10\;\mbox{GHz}$, so
that the RF drive looks constant to the SET, and the dc-SET equations
can be used.

\begin{figure}[b]
\centering
\includegraphics[width=0.8in]{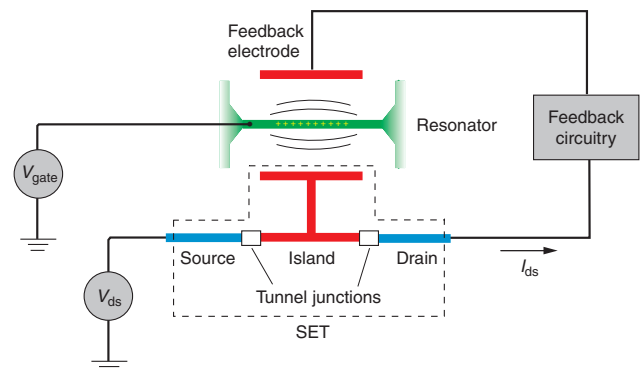}
\caption{A schematic of the resonator, measuring and feedback apparatus.  As
the resonator moves closer to the SET, the current flowing through the
SET changes, and that information is then used to generate a feedback
voltage applied to an actuating gate.}
\label{figSET}
\end{figure}

We will use a quantum mechanical model of the measurement and feedback
process, but discuss how in this case, such a description is
equivalent to a classical measurement of a noisy classical system.
Thus, this article is intended for both experimentalists familiar with
classical descriptions of noise in systems, as well as quantum
measurement theorists.

Rather than performing a microscopic analysis of the measurement
process in terms of the interaction of the SET and the resonator, we
start by introducing equations which describe the continuous
observation of a quantum observable, and show how this includes the
shot noise and back-action, these being the key sources of noise in a
continuous quantum measurement. This description can then be tailored
to the case of a measurement with a SET by choosing the parameters so
that the noise sources match those calculated in microscopic noise
analyses which have been performed for the
SET~\cite{Korotkov94,Zhang}.

A treatment of the continuous quantum measurement of a two-state
system using a SET has been carried out by Korotkov~\cite{Korotkov01},
using what might be referred to as a partially microscopic approach.
The equations we use here may be derived by replacing the two-state
observable in those equations by the resonator position.  A full
analysis, along the lines of those performed for quantum optical
systems~\cite{Wiseman93,Milburn94}, can also be expected to produce
the same equations under reasonable approximations. The form of these
equations is determined by how information is obtained, and not by the
specific implementation, which explains why the form of the equations
is similar in optical position measurements and position measurement
using SETs.  If the measurement is of a physical observable, and the
resulting error about the expectation value of that observable in a
short time interval $\Delta t$ is Gaussian, then the most
straightforward implementation of that measurement process has the
form used here.

In the next section we introduce the equations that describe a
continuous measurement process, derive the form of the resulting
noise, and give the equivalent classical model.  We then discuss how
this model can be applied to position measurement using a SET, and
compare our formulae to those derived using a semi-classical treatment
of the SET~\cite{Korotkov94,Zhang} in order to express our results in
terms of experimental parameters.  In Section III we discuss the
implementation of a feedback algorithm and calculate the minimum
achievable temperature in terms of physical parameters.  We then
calculate estimates of realistic achievable temperatures for an an
experimentally realizable sample system in Section IV, and finally
conclude with a summary of the results obtained.

\section{Continuous Quantum Measurement of Position}

Given a quantum system whose state is specified by the density matrix
$\rho$, and whose evolution is determined by the Hamiltonian $H$, then
a continuous measurement of the observable $O$ of that system, which
provides the continuous output results (measurement record), 
\bq 
  dr = \langle O \rangle dt + \frac{1}{\sqrt{8 k}} dW,
      \label{Eq1}
\eequ 
induces the following evolution of the
system~\cite{Caves87,Jacobs98,Doherty99}  
\begin{eqnarray}
d\rho & = & -i[H,\rho]dt-k[O,[O,\rho]]dt \nn \\
      &   & +\ \sqrt{2 k}(O\rho+\rho O - 2\langle O \rangle)dW.
      \label{Eq2}
\end{eqnarray}
Here $k$ is proportional to the measurement strength and $dW$ is a
Weiner process.  The noise contained in the measurement record is a
necessary result of the fact that only a finite amount of information
is obtained regarding the observable $O$ in a finite time.  This
direct noise on the record is called the {\em shot noise}.  However,
this is not the only noise resulting from the measurement process.  As
a result of Heisenberg's uncertainty relation, information about one
observable makes other observables less certain.  Due to the dynamics,
the uncertainty (noise) in these observables can feed into the
observable being measured.  This source of noise is referred to as
{\em back-action}.  If the Hamiltonian is such that the increased
uncertainty is not fed back into the observable being measured, then
the measurement is referred to as `back-action evading.'

Now let us examine the case of a position measurement on a harmonic
oscillator.  To do this, we set $O=x$, and the Hamiltonian becomes
\begin{equation}
H = \frac{p^{2}}{2m} + \frac{1}{2}m \omega_0^{2} x^{2},
\label{ham}
\end{equation}
where $m$ is the mass of the particle, $\omega_0$ is the (angular)
frequency of the oscillation, and $x$ and $p$ are the position and
momentum operators, respectively. To make our model sufficiently
realistic, we need to include two more sources of noise: the first is
the intrinsic thermal noise of the harmonic oscillator, and the second
is the possibility that the oscillator may be driven by white noise
over and above that required by Heisenberg's uncertainty principle
(excess `technical noise').

The second of these is easily included by adding a term $-\beta
[x,[x,\rho]]$ to the equation of motion of $\rho$; this describes a
noise term identical to the one caused by the back-action, but without
the corresponding dynamics of $\rho$ associated with obtaining a
measurement result which causes the back-action. It is equivalent to
adding a term linear in $x$ to the Hamiltonian (\ref{ham}) multiplied
by white noise.

The inclusion of thermal fluctuations is only a little more involved,
and can be achieved by coupling the oscillator to a thermal bath.  In
our case the effect of the thermal bath may be included by adding the
`standard Brownian motion master equation' (SBMME)~\cite{QNoise} to
our equation of motion for $\rho$:
\begin{eqnarray}
d\rho & = & -\frac{i}{\hbar}[H,\rho]dt -\ \frac{i \Gamma}{2\hbar} [ x,
\{ p,\rho \}_+ ] dt 
\nonumber \\ 
& & - \left( k + \beta + \frac{m \omega_0 \Gamma}{2 \hbar} \coth
\frac{\hbar \omega_0}{2 k_B T} \right) \mbox{\bf [} x,[x,\rho]
\mbox{\bf ]} dt \nn \\ 
& & +\ \sqrt{2 k}( x\rho + \rho x - 2\langle x \rangle \rho )dW 
\end{eqnarray}
where $\Gamma = \frac{\omega}{Q}$, $Q$ being the quality factor of the
resonator.  The two terms proportional to $\Gamma$ are due to the
inclusion of the SBMME, the first representing dissipation due to the
reservoir while the second is a diffusion term due to environmental
fluctuations. Here we are using an approximate form of the SBMME
appropriate for the weak coupling regime (small $\Gamma$, large $Q$)
but covering all ranges of temperatures~\cite{CCR}. Since the
nanomechanical resonators we consider all have large values of $Q$,
the weak coupling requirement is easily satisfied. The temperature
dependence of the diffusion coefficient is given by $\coth(\hbar
\omega_0/2 k_B T)$ so that the diffusion does not vanish as
$k_BT\rightarrow 0$: this correctly accounts for the existence of
quantum vacuum fluctuations which exist even at zero temperature. In
the absence of a rigorous characterization of the dissipation channels
of nanomechanical systems there is as yet no need to include a more
sophisticated description of SBMME environmental effects~\cite{HuVit}.
Phenomenological corrections to the SBMME such as temperature
dependence of $\Gamma$ can be added if needed, but these are not
significant effects in the high-$Q$ regime.

We also need to include in our model the possibility that there is
noise driving the oscillator which is correlated with the noise on the
measurement record (the shot noise). This can happen if the noisy
behavior of the oscillator explicitly causes some of the noise in the
measurement apparatus, or vice versa.  In this situation, the
measurement record contains more information about the oscillator
position, so when it comes to adding feedback, we are able to cool the
oscillator further than would otherwise be expected.  In
Eq.~(\ref{Eq2}) the noise driving the oscillator is purely the quantum
back-action.  It may appear from Eqs.~(\ref{Eq1}) and (\ref{Eq2}) that
the quantum back-action is correlated with the shot noise due to the
fact that the same noise term ($dW$) appears in both equations.
However, this is not the case. The term proportional to $dW$ which
appears in the equation for $\rho$ describes the random way in which
the measurement changes the observers state of knowledge about the
system. Thus, in general, this noise term {\em decreases} the entropy
of $\rho$.  The back-action noise, which is driving the oscillator and
consequently {\em increasing} the entropy of $\rho$, is described by the
term proportional to $k$. The quantum back-action is, in fact,
completely uncorrelated with the shot noise.

To drive the oscillator with a random force, one applies the
Hamiltonian $\hbar\xi(t)x$, where $\xi(t)$ is the magnitude of the
random force. We can choose $\xi(t)$ to be correlated with the shot
noise, with correlation coefficient $\kappa$, by setting
\begin{equation}
  d\xi=\sqrt{2\alpha}(\sqrt{\kappa}dW + \sqrt{1-\kappa}dV), 
\end{equation}
where $dV$is a Wiener noise uncorrelated with $dW$. The resulting
spectral density of $\xi(t)$ is $\alpha$, so that
$\langle\xi(t)\xi(t')\rangle = \alpha\delta(t-t')$. The Stratonovich
equation which describes the driving by $\xi(t)$ is
\begin{equation}
 \label{xiEq1}
|\dot{\psi}\rangle = -i\xi(t)x|\psi\rangle ,
\end{equation}
and converting this to an Ito equation gives
\begin{equation}
  d|\psi\rangle = -i \sqrt{2\alpha}x|\psi\rangle d\xi - \alpha x^2 |\psi\rangle dt .
\end{equation}
Converting the Ito equation further to an equation for $\rho$ one obtains
\begin{equation}
  d\rho = -\alpha[x,[x,\rho]]dt - i\sqrt{2\alpha}[x,\rho]d\xi .
\end{equation}
Since the observer has access to $dW$, but not to $dV$, she must
average over $dV$, and this gives
\begin{equation}
d\rho= -\alpha[x,[x,\rho]]dt - i\sqrt{2\kappa\alpha}[x,\rho]dW.
\end{equation}
If we allow part of the excess noise given by $\beta$ in our model to
be due to driving by the shot noise $dW$ (that is, this noise is
correlated with the shot noise $dW$ with correlation coefficient
$\kappa$) then the equation of motion for the system becomes
\begin{eqnarray}
\label{masterEqn}
d\rho & = & -\frac{i}{\hbar}[H,\rho]dt-\
\frac{i\Gamma}{2\hbar}[x,\{p,\rho \}_+ ] dt \nonumber \\
& & -\ \left( k + \beta + \frac{m \omega_0 \Gamma}{2 \hbar} \coth
\frac{\hbar \omega_0}{2 k_B T}\right) [x,[x,\rho]]dt \nn \\ 
& & -\ i\sqrt{2\kappa\beta}[x,\rho]dW \nn \\ 
& & +\ \sqrt{2 k}(x\rho+\rho x -2 \langle x \rangle \rho)dW
\end{eqnarray}
This completes our quantum mechanical description of a resonator under continuous
observation.

Now that we have an equation that includes all the relevant noise
terms, the noise spectrum of the measurement record can be obtained:
\begin{eqnarray}
S(\omega) & = & \frac{1}{8k} + \left( k + \beta + \frac{m \omega_0
\Gamma}{2 \hbar} \coth \frac{\hbar \omega_0}{2 k_B T}\right)\nn\\ 
& & \times\ \frac{2(\hbar/m)^2}{\Gamma^2 \omega^2 +
(\omega^2-\omega_0^2)^2} 
\label{spectrum}
\end{eqnarray}
The first term is the shot noise, which is white, the term
proportional to $k$ is the quantum back-action, the term proportional
to $\Gamma$ is the effect of the noise from the resonator's thermal
environment, and the term proportional to $\beta$ gives any excess
noise over and above the necessary quantum back-action.  Note that the
last three terms all have the same form as a function of $\omega$.
This is because they are all white noises filtered through the
harmonic oscillator spectral response function.

While our treatment so far has been fully quantum mechanical, it is
worth noting that a completely classical model of a measured, damped
oscillator will reproduce the dynamics of this measured quantum
system, so long as the initial density matrix is Gaussian in $x$ and
$p$~\cite{Doherty99}. Thus, one can understand the behavior of the
oscillator in terms of classical noise and a classical measurement
process. The equations of motion for the position, $x_{\ms c}$, and
momentum $p_{\ms c}$ of this equivalent classical oscillator are
\begin{eqnarray}
  dx_{\ms c} & = & \frac{1}{m} p_{\ms c} dt \\
  dp_{\ms c} & = & -m\omega_0^2 x_{\ms c} dt - \Gamma p_{\ms c} dt + \hbar\sqrt{2k}dY_{\ms c} 
             + \hbar\sqrt{2\beta} dV_{\ms c} \nn \\
       &   & + \sqrt{m \hbar \omega_0 \Gamma \coth \frac{\hbar \omega_0}{2 k_B T}} dU_{\ms c}
\end{eqnarray}
where $dY_{\ms c}$, $dV_{\ms c}$ and $dU_{\ms c}$ are each zero-mean
Gaussian white noise, and mutually uncorrelated. The position of the
oscillator is then observed by a continuous classical measurement,
which generates the output record
\begin{equation}
  dr_{\ms c} = x_{\ms c} dt + \frac{1}{8 k} dZ_{\ms c} ,
\end{equation}
and where $dZ_{\ms c}$ is zero-mean Gaussian white noise, uncorrelated
with $dY_{\ms c}$. The noise term $dY_{\ms c}$ is what is required in
the classical model to correctly include the back-action of the
quantum measurement process.  It is now explicit that this noise is
uncorrelated with the shot noise on the measurement, $dZ_{\ms c}$.

In the classical case, the observer's state of knowledge about the
oscillator is given by a joint probability density over $x_{\ms c}$
and $p_{\ms c}$. This probability density is the classical equivalent
of the density matrix $\rho$. So long as the initial probability
density is Gaussian, it remains Gaussian as time passes, and as a
result the observer's full state of knowledge may be represented by
merely 5 variables: the mean position and momentum, $\langle x_{\ms c}
\rangle$ and $\langle p_{\ms c} \rangle$, and the variances and
covariance, given by
\begin{eqnarray}
\sigma^2_x    & = & \langle x_{\ms c}^2 \rangle - \langle x_{\ms c}\rangle^2,\\ 
\sigma^2_p    & = & \langle p_{\ms c}^2 \rangle - \langle p_{\ms c}\rangle^2,\\ 
\sigma^2_{xp} & = & \langle x_{\ms c} p_{\ms c} \rangle - \langle x_{\ms c}\rangle
\langle p_{\ms c}\rangle .
\end{eqnarray}
It is the means $\langle x_{\ms c} \rangle$ and $\langle p_{\ms c}
\rangle$ (being the observer's best estimates of the value of $x_{\ms
  c}$ and $p_{\ms c}$) which are the classical equivalents of the
quantum expectation values $\langle x \rangle$ and $\langle p
\rangle$. It turns out that if one writes the classical measurement
record as
\begin{equation}
 dr_{\ms c} =\langle x_{\ms c} \rangle dt + \frac{1}{8 k} dW_{\ms c} ,
\end{equation}
then $dW_{\ms c}$ is zero-mean Gaussian white noise~\cite{Jacobs},
uncorrelated with $dZ_{\ms c}$. The classical model is then equivalent
to the quantum model if we equate $dW_{\ms c}$ with the quantum
measurement noise, $dW$, and correlate $dV_{\ms c}$ with $dW_{\ms c}$,
so that $\langle V_{\ms c}(t)dW_{\ms c}(t') \rangle =
\kappa\delta(t-t')$.

\section{Continuous Measurement with a Single Electron Transistor}
\label{SETmeas}

Having obtained a model which is sufficiently general to encompass the
dynamics of a resonator monitored by a SET, we need to express the
theoretical parameters $k$, $\beta$, and $\kappa$ in terms of the
actual experimental parameters of the SET. Since it is by measuring
current through the SET that we measure the resonator position, it is
the spectral density of this current which determines the shot noise
of the measurement. The back-action from the measurement is due to the
action of the SET on the resonator, which is the force that the
resonator feels from the charge on the SET island. As a result the
back-action noise, $\beta$, can be calculated from the spectral
density of the charge fluctuations on the SET island, and hence
$\kappa$ is determined by the correlation between the current and the
island charge fluctuations.

However, the dynamics of the SET are sufficiently complex that
analytic results for these spectra have as yet only been obtained for
certain parameter regimes.  These calculations have been performed by
Zhang and Blencowe~\cite{Zhang}, using previous results of
Korotkov~\cite{Korotkov94}. The technique used is to approximate the
dynamics of the electron tunneling on and off the SET island by a
classical master equation. That is, the electrons are assumed to
tunnel independently across each of the junctions, with certain rates
(the rates being obtained using a perturbative quantum calculation).
This ignores the possibility that electrons will tunnel coherently
across both junctions simultaneously, a quantum effect referred to as
{\em cotunneling}.

However, it is important to note that the above ``classical'' method
for calculating the charge fluctuations does not include the quantum
back-action noise. This can be seen from the following
argument~\cite{HWpriv}. In the classical treatment, since the 
fluctuating force on the resonator is due to the electrons jumping on
and off the island, in principle the time history of this force can be
known by detecting the electrons flowing in the circuit. In principle
then, the effect of the noise can be known, and if desired, undone. As
a result it cannot include the quantum back-action, since this cannot,
even in principle, be undone.

The approach we will take here is to use the classical calculation of
the force noise on the resonator, which determines $\beta$, and add to
it the necessary back-action as required by quantum mechanics, which
is given by $k$. In doing this we note that the ratio $k/\beta$
provides a diagnostic tool for determining when the classical
calculation breaks down; if $k/\beta \ll 1$ is not satisfied, then the
classical calculation no longer provides a good estimate of the total
force noise on the resonator. Thus it should be noted that if $k/\beta
\gtrsim 1$, then the classical calculation cannot be relied upon. That
is, it is possible in this case that the total noise on the resonator
is significantly larger than our estimate $k + \beta$, due to quantum
contributions not taken into account in the classical calculation.

We find that in the regions of best cooling, which we explore in the
following, $k$ is larger than $\beta$ (although near-optimal cooling
can be obtained with $k\leq \beta$, and in particular we will give as
an example results for $k=\beta/4$). Hence our calculations should be
regarded as estimates of the performance of the feedback algorithm,
rather than exact results. We note, however, that a more sophisticated
analysis using the diagrammatic techniques developed by Schoeller and
Sch\"{o}n~\cite{Schoeller} might provide analytic, or semi-analytic
results for the parameter regime of most interest for quantum
measurement and control, and therefore may provide a method for more
accurate calculations.

The spectral densities given by the classical calculation are derived
in Appendix~\ref{Spectra}. Approximations which are used in the
derivation are detailed there, and come primarily from Zhang and
Blencowe~\cite{Zhang}. The noise spectrum of the displacement of the
resonator due to the shot noise of the SET current is
\begin{eqnarray}
    S_{X}^{I}&=& \frac{S_{\mbox{\scriptsize I}}(\omega)}{(dI_{\mbox{\scriptsize ds}}/dx)^2},
\label{Zhang_shot}
\end{eqnarray}
where $S_I(\omega)$ is the spectral density of the shot noise, given
in Eq.~(\ref{SI}), and $I$ is the current through the SET, given in
Eq.~(\ref{current}). The dependence of the current on the
displacement of the resonator comes from its dependence on the gate
capacitance, which can be approximated by
\begin{eqnarray}
    C_{\ms{g}} \approx C_{g0}(1-\frac{x}{d}).
\label{Cg_x}
\end{eqnarray}
The shot noise, $S_I(\omega)$ is, to a very good approximation, frequency 
independent, as required by our quantum measurement model. Thus
\begin{eqnarray}
    \frac{1}{8k} = S_{X}^{I}|_{\omega=0} = \left. \frac{S_{\mbox{\scriptsize I}}(\omega)}{(dI_{\mbox{\scriptsize ds}}/dx)^2} \right|_{\omega=0} .
\end{eqnarray}
The spectral density of the classical part of the displacement noise
due to the fluctuating force on the resonator is
\begin{eqnarray}
    S_{X}^{F}(\omega)&=& \frac{S_{\mbox{\scriptsize F}}(\omega)/m^2}{\Gamma^2 \omega^2 + (\omega^2-\omega_0^2)^2},  
\label{Zhang_force}
\end{eqnarray}
where $S_F(\omega)$ is the spectral density of the fluctuating force
given in Eq.~(\ref{SF}). Since, once again, $S_F(\omega)$ is
effectively frequency independent, we have
\begin{equation}
\beta = \left. \frac{S_{\mbox{\scriptsize F}}}{2 \hbar^2} \right|_{\omega=0} .
\end{equation}
The correlation coefficient, $\kappa$, between the shot noise and the
excess back action is therefore simply the correlation, $C$, between
$S_{\mbox{\scriptsize I}}$ and $S_{\mbox{\scriptsize F}}$, which is
given in Eq.~(\ref{C2}).

\section{Feedback control}

We wish to cool the dynamics of the resonator by using the information
obtained continuously about the state of the resonator to direct a
time-dependent external force. Such a force may be applied, for
example, by passing current through the resonator and immersing it in
a magnetic field.  It can also be applied by placing an actuating gate
near the resonator, and varying the potential difference between the
charged resonator and the actuating gate.

In this case the results of modern optimal control theory apply, since
the dynamics of the resonator are equivalent to that of a classical
oscillator driven by Gaussian noise, so long as we restrict ourselves
to a linear external force~\cite{Doherty99,Doherty00}. This allows us
to obtain the optimal feedback algorithm in a straightforward manner.
Choosing the minimization of the energy of the resonator as the
feedback objective it turns out that as long as the force we apply is
sufficiently large, this force should be chosen to be~\cite{Doherty99}
\begin{equation}
  F = -\gamma (m \omega_0 \langle x \rangle + \langle p \rangle) ,
\end{equation}
where $\gamma$ is a rate constant which determines the overall
strength of the force. This equation gives optimal performance so long 
as $\gamma \gg \omega_0$, which is within reach of current experiments, 
as detailed below.

To calculate the average energy of the controlled resonator, we first
need the equations of motion for the mean values in the continually
observed and controlled case, which are
\begin{eqnarray}
\label{dexpectx}
  d\la x \ra & = & \frac{\la p \ra}{m}dt - \Gamma \la x \ra dt + 2
  \sqrt{2 k}\sigma^2_x dW \\ 
  d\la p \ra & = & -m \omega^2 \la x \ra dt - 2 \Gamma \la p \ra dt -
  \gamma(m \omega \la x \ra + \la p \ra)dt \nn \\  
\label{dexpectp}
 & & +\ \sqrt{2 \kappa \beta} \hbar dW + 2 \sqrt{2 k}\sigma^2_{xp} dW, 
\end{eqnarray}
and for the covariances,
\begin{eqnarray}
  \dot{\sigma^2_x}    & = & \frac{2}{m}\sigma^2_{xp} -8 k
  (\sigma^2_x)^2 , \\ 
  \dot{\sigma^2_p}    & = & -2 m \omega^2 \sigma^2_{xp}  -8 k
  (\sigma^2_{xp})^2 - 2 \Gamma \sigma^2_p + 2\hbar^2 k \nn \\ 
& &  + 2 \hbar^2 \left[ (1-\kappa)\beta+\frac{m \omega_0 \Gamma}{2 \hbar} \coth
  \frac{\hbar \omega_0}{2 k_B T} \right] \\ 
  \dot{\sigma^2_{xp}} & = & \frac{\sigma^2_p}{m} -m \omega^2
  \sigma^2_x - \frac{\Gamma}{2} \sigma^2_{xp} \nonumber - 8 k
  \sigma^2_x \sigma^2_{xp}  \\ 
  & & -\ 4 \sqrt{\kappa \beta k}\hbar \sigma^2_x.
\end{eqnarray}
In these equations, $\sigma_x^2$ and $\sigma_p^2$ are the variances in
position and momentum, respectively, and
\begin{equation}
\sigma^2_{xp} = \frac{1}{2}\la xp+px\ra - \la x \ra \la p \ra
\end{equation}
is the symmetrized covariance. This system of equations is exactly
equivalent to Eq.~(\ref{masterEqn}) as long as the initial state is
Gaussian. In order to solve this set of equations most easily, we make
what we call the truncated Gaussian approximation. We assume that the
feedback rate $\gamma$ is much larger than the system's small
intrinsic damping $\Gamma$, and we therefore drop all terms
proportional to $\Gamma$ from the above equations. This approximation
is easily justified for current experiments.

The steady state solutions to these equations are
\begin{eqnarray}
\sigma^2_x & = & \frac{\sqrt{2} \omega}{8 k}\sqrt{\Lambda} , \\
\sigma^2_p & = & \frac{\sqrt {2} m^2 \omega^3}{8 k}[\sqrt{\Lambda} +
\Lambda^{3/2}] \nonumber \\ 
           & & +\ \frac{4 \sqrt{2} \hbar m \omega}{k} \sqrt{\kappa
           \beta k}\sqrt{\Lambda}, \\ 
\sigma^2_{xp} & = & \frac{m \omega^2}{8 k}\Lambda ,
\end{eqnarray}
where
\bqa
&&\Lambda +1 = \nonumber\\
&& \left[1 + 16 \frac{k\hbar^2 \left\{k+(1-\kappa)\beta+\frac{m
\omega_0 \Gamma}{2 \hbar} \coth \frac{\hbar \omega_0}{2 k_B
T})\right\}}{m^2 \omega^4}\right]^{1/2}.  
\label{pieq}
\eqa
In the limit of both large and small values of $k$, $\Lambda\sim k$.

Using the steady state solutions for the variances $\sigma^2_x$ and
$\sigma^2_{xp}$ in Eqs.~(\ref{dexpectx}) and (\ref{dexpectp}), and
calculating the variances of $\la x \ra$ and $\la p \ra$, we obtain 
\bqa
\sigma^2_{\la x \ra} & = & \frac{\omega(\gamma^2 + \gamma \omega +
\omega^2)}{8 k \gamma (\omega + \gamma)}\Lambda  + \frac{\sqrt{2}
\omega^2}{8 k (\omega+\gamma))}\Lambda^{3/2} \nn \\ 
& & +\ \frac{\omega^3}{16 k \gamma (\omega + \gamma)}\Lambda^2 +
\frac{\kappa \beta \hbar^2}{m^2 \omega \gamma (\omega + \gamma))} \nn
\\ 
& & +\ \frac{\hbar \sqrt{\kappa \beta k}}{2 m (\omega +
\gamma)}\left[\frac{\sqrt{2 \Lambda}}{k} + \frac{\omega \Lambda}{2
\gamma}\right]\\  
\sigma^2_{\la p \ra} & = & \frac{m^2 \omega^3(\omega+\gamma)}{8 k
\gamma} \Lambda + \frac {m^2 \omega^4}{16 k \gamma}\Lambda^2 +
\frac{\kappa \beta \hbar^2}{\gamma} \nn \\ 
& & +\ \frac{m \omega^2 \hbar \sqrt{\kappa \beta k}}{4 k
\gamma}\Lambda. 
\eqa

The average steady-state energy of the oscillator under observation
and feedback is the sum of the intrinsic variances of the Gaussian
steady state and the variances in the observer's measurement record.
Thus
\begin{eqnarray}
E & = & \frac{1}{2} m \omega^2 (\sigma^2_x + \sigma^2_{\la x \ra}) +
\frac{\sigma^2_p + \sigma^2_{\la p \ra}}{2 m}\\ 
& = & \frac{m \omega^3}{8 k}\left[\sqrt{2 \Lambda} + \Lambda +
\frac{\sqrt{2}}{2}\Lambda^{3/2} + \frac{\omega}{4
\gamma}\Lambda^2\right] + \frac{\kappa \beta \hbar^2}{2 m \gamma} \nn \\ 
\label{energy} 
& & +\ \frac{\hbar \omega \sqrt{\kappa \beta k}}{4 k}[\sqrt{2 \Lambda}
+ \frac{\omega}{2 \gamma}\Lambda]. 
\label{ee}
\end{eqnarray}
Here we have used the simplifying assumption $\gamma \gg \omega$, 
since this is inherent in the optimal control condition.

It is clear from Eqs.~(\ref{pieq}) and (\ref{ee}) that reducing the
background temperature allows for lower final temperatures. Extremely
low values of $k$ lead to heating as can be seen from the fact that
$\Lambda\sim k$.  For large $k$ (corresponding to large gate voltage),
the increased sensitivity of the measurement cancels the increased
disturbance due to the measurement, with the result that the minimal
temperature levels off as $k$ is increased.

\section{Estimates for achievable temperatures}

Current refrigeration technology allows experiments on nanomechanical
resonators to be performed at temperatures of about $100\;\mbox{mK}$.
It is therefore sensible to assume that the feedback algorithm will be
applied to a device which is initially at this temperature. In such
experiments the resonators typically have fundamental frequencies in
the range $f_0 = 1-100\;\mbox{MHz}$. As our example system we take a
realistic resonator with $f_0 = 12\;\mbox{MHz}$, which is
$6~\mu\mbox{m}$ in length, $50\;\mbox{nm}$ wide, and $150\;\mbox{nm}$
thick. We restrict ourselves to relatively low frequencies because of
the limits of feedback circuitry, which we estimate can easily operate
at $50\;\mbox{MHz}$. The effective mass of such a resonator is roughly
$10^{-16}\;\mbox{kg}$. An achievable quality factor, $Q$, is on the
order of $10^4$.

\begin{figure}
\centering
%\leavevmode\rotatebox{-90}{\includegraphics[width=2.4in,height=3.3in]{plotbyT.epsc}}
\leavevmode\includegraphics[width=3.4in]{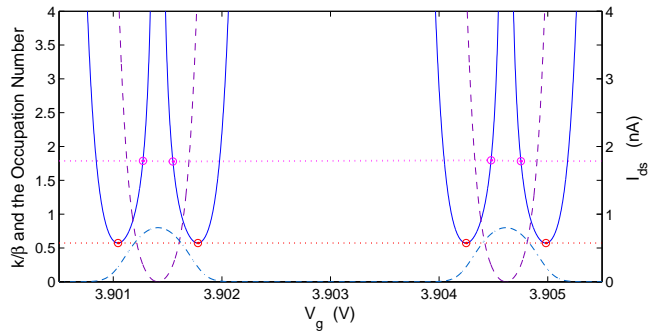}
\caption{The steady-state average occupation number, $\langle N \rangle$,
  as a function of the gate voltage (solid line), plotted
  along with the ratio $k/\beta$ (dashed line), and the drain-source
  current, $I_{\ms ds}$ (dot-dash line). The lower dotted line gives
  the minima of $\langle N \rangle$, and the upper dotted line gives
  the value of $\langle N \rangle$ when $k/\beta = 1/4.$}
\label{Tzoom}
\end{figure}

Realistic values for the resistances and capacitances of the junctions
of a SET which would be used to monitor the resonator are $R_1 = R_2 =
50\;\mbox{k}\Omega$, and $C_1 = C_2 = 100\;\mbox{aF}$, and we place it
$d \sim100\;\mbox{nm}$ from the resonator. We estimate that the
capacitance between the gate of the SET and the resonator will be
roughly $C_{\ms{g}} = 50\;\mbox{aF}$, so that $C_\Sigma =
250\;\mbox{aF}$ ($C_\Sigma = 2 C_j + C_{\ms{g}}$). It is important to
note that the analysis we use in the appendix to obtain the noise
spectra is only a good approximation in certain parameter regimes. In
particular, we require that $V_{\ms{ds}}$, being the drain-source
voltage across the SET, satisfies $V_{\ms{ds}} \ll e/C_\Sigma$.

To apply the feedback force, we place the resonator $100\;\mbox{nm}$
from the actuating gate, and allow the controller to vary the voltage
difference between the gate and the resonator between $-4\;\mbox{V}$
to $4\;\mbox{V}$.  The capacitance of this arrangement is about
$50\;\mbox{aF}$, so the maximum force that can be applied to the
resonator is of the order of $10^{-8}~N$.  This corresponds to $\gamma
\approx 1.08 \times 10^{13}\;\mbox{s}^{-1}$, which is much larger than
$\omega$ and $\Gamma$, as required by the optimal control condition
and truncated Gaussian approximation used in the previous section.

Before giving theoretical estimates for the achievable steady-state
temperature (or equivalently, the steady-state average occupation
number of the oscillator, $\langle N\rangle = \langle a^\dagger a
\rangle$), we need to explain two subtleties which affect the
presentation of our results. When one examines the dependence of the
steady-state $\langle N \rangle$ on the gate voltage,
once finds that it oscillates very rapidly, with minima occurring in
closely spaced pairs. Since $V_{\ms g}$ is experimentally easy to
tune, all else being equal it would make sense simply to plot these
minima and ignore the complex structure. However, the locations of the
minima are such that $k/\beta$ can be large, and as discussed in
section~\ref{SETmeas}, our results are more trustworthy the smaller
$k/\beta$. The situation is shown in detail in Figure~\ref{Tzoom}, in
which we display, as a function of $V_{\ms g}$, two pairs of the
$\langle N \rangle$ minima, as well as $k/\beta$ and the current
$I_{\ms{ds}}$. When presenting results in what follows, we will plot
both the minima with respect to $V_{\ms g}$, and the temperature
which results if we demand that $k/\beta \leq 1/4$. For clarity these
points are also displayed in Figure~\ref{Tzoom}.

As an example of the relative magnitudes of the various noise sources
at the minima displayed in Figure~\ref{Tzoom}, if we set the
drain-source voltage at $V_{\ms{ds}} = e/(4C_\Sigma) =
0.16\;\mbox{mV}$ and the gate voltage at $V_{\ms{g}} \sim
1\;\mbox{V}$, then the noise sources are:
\begin{eqnarray}
    & & \beta = 4.47\times10^{29}\ \mbox{m}^{-2} \mbox{s}^{-1} , \\
    & & k = 4.2 \;\beta , \\
    & & \frac{m \omega_0 \Gamma}{2 \hbar} \coth \frac{\hbar \omega_0}{2 k_B T} = 209 \;\beta
\end{eqnarray}
and the correlation coefficient is $\kappa = 0.633$.

Using the above parameter values to calculate the minimum achievable
temperature, we find that $\Lambda = 4.95 \times10^{-5}$, and $T =
2.06\;\mbox{mK}$.  This corresponds to an energy of about
$E_{\mbox{\scriptsize ss}} = 2.85\times10^{-26}\;\mbox{J}$, and an
average occupation number $\la N \ra = 3.09$.  While this is very
encouraging, ideally one wants to cool below the energy of the first
excited state, and we now examine what is required to do this.

\begin{figure}
\centering
\leavevmode\includegraphics[width=3.3in]{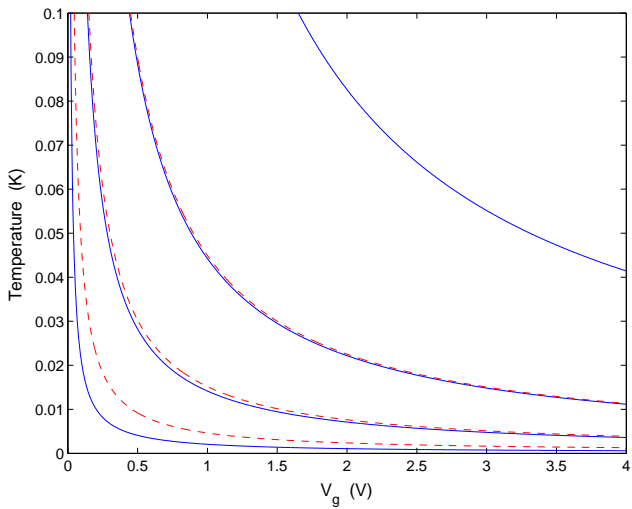}
\caption{Estimates for minimum achievable temperatures as a function 
of gate voltage for a range of initial temperatures. The dotted lines give the
  minimum temperature under the additional restriction that $\beta=4k$. From top to
bottom, the initial temperatures are $2\;\mbox{K}$, $1\;\mbox{K}$,
$500\;\mbox{mK}$, and $100\;\mbox{mK}$.}
\label{plotbyT}
\end{figure}

\begin{figure}
\centering
\leavevmode\includegraphics[width=3.3in]{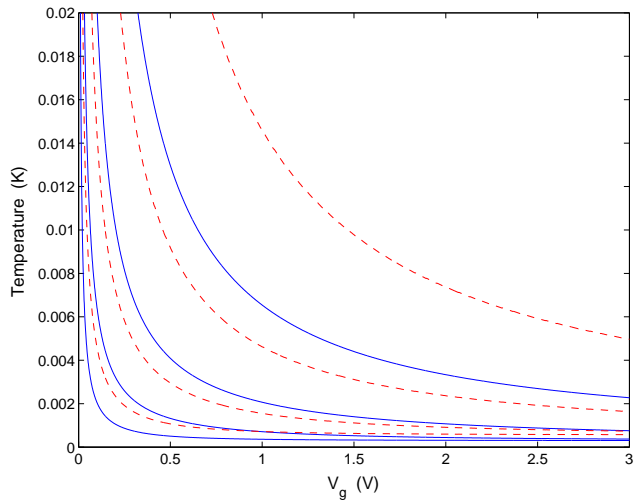}
\caption{Estimates for minimum achievable temperatures as a function
  of gate voltage for a range of resonator quality factors and an
  initial temperature of $100\;\mbox{mK}$. The dotted lines give the
  minimum temperature under the additional restriction that $\beta=4k$. From top
  to bottom, the quality factors are $10^3$, $10^4$, $10^5$, and
  $10^6$.  A quality factor of $10^4$ is achievable with current 
  technology.}
\label{plotbyQ}
\end{figure}

While classically an increase in measurement strength would
automatically lead to improved tracking of the resonator, and
therefore more efficient cooling, quantum mechanically the situation
is more complex due to the fact that more precise measurement also
leads to increased heating due to back-action. Nevertheless, In the
present case one finds that the increased sensitivity of the
measurement with increasing measurement strength effectively cancels
this heating, and as a result larger $V_{\ms{g}}$ corresponds to
better cooling.  However, after a sharp increase in cooling with
increasing $V_{\ms{g}}$, the minimal temperature levels off, so
greater $V_{\ms{g}}$ no longer provides much benefit.  In addition, at
some $V_{\ms{g}}$ snap-in is likely to occur as the force between the
SET gate and the resonator becomes too strong. This voltage, in our
example system, is estimated to be roughly $4\;\mbox{V}$.  As a
result, we limit ourselves to $V_{\ms{g}} \leq 4\;\mbox{V}$.  At
$V_{\ms{g}} = 4\;\mbox{V}$ the steady state energy $E = 8.37
\times10^{-27}\;\mbox{J}$, which is below the energy of the first
excited state. This corresponds to $T = 0.61\;\mbox{mK}$, and $\la N
\ra = 0.555$. Thus, if the energy were to be measured directly,
immediately after turning off the feedback, energy jumps as a
signature of quantum behavior may well be observable.  As an
indication of the return from increasing the gate voltage, the steady
state energy is $E = 1.48\times10^{-26}\;\mbox{J}$ for $V_{\ms{g}}
\approx 2\;\mbox{V}$, which corresponds to $\la N \ra = 1.36$.

In Figure~\ref{plotbyT} we plot the theoretical estimates for the
achievable steady-state temperature as a function of $V_{\ms{g}}$ for
a range of starting temperatures.  The solid lines correspond to the
absolute minima, and the dotted lines to the minima under the
restriction that $k \leq \beta/4$. Of particular interest is the fact
that for a starting temperature of $2\;\mbox{K}$ (ie.\ with pumped
liquid He), we obtain minimum temperatures in the range of
$50\;\mbox{mK}$. Thus, even for an initial temperature of
$2\;\mbox{K}$, feedback cooling might well be able to compete with
dilution refrigerators.  If the resonator is first cooled in a
dilution refrigerator, and then feedback cooled, then the
semi-classical theory predicts achievable temperatures below
$1\;\mbox{mK}$.  In Figure~\ref{plotbyQ} we plot the dependence of the
minimum temperature on $V_{\ms{g}}$ for a range of quality factors,
which shows that final temperatures may be increased somewhat by
increasing the $Q$.

\section{Discussion and Conclusion}

The results obtained above are consistent with heuristic arguments.
The response of cooling to the measurement strength is as expected:
for very weak continuous measurements, we do not learn enough about
the state of the system to cool it effectively, and can in fact heat
the system due to acting on our poor information.  For very strong
continuous measurements, we gain sensitivity, but inject more quantum
back-action, and approach a minimum only asymptotically. The range of
improvement is limited, however, and beyond a few volts, the benefits
may not warrant the additional effort.

Higher drain-source voltages provide a larger signal-to-noise ratio,
and therefore improve cooling.  However, since we do not know exactly
how our approximations will fail as $V_{\mbox{\scriptsize ds}}$
approaches $e/C_\Sigma$, and we lack a complete theory of the SET once
more than 2 island states need to be taken into account, we have
chosen to stay below that limit.

We have made a few additional simplifying assumptions, as a way to
indicate a goal, rather than an immediately achievable experimental
realization.  First, we have assumed a perfectly efficient (and
infinite bandwidth) measurement -- that is, that no electron passes
the detector without being detected.  While detection efficiency is not
as much of a problem here as in optical experiments, detectors will
necessarily be inefficient to some extent. Second, we have assumed
perfect, noiseless feedback.  In reality, the actuating gate applying
the feedback voltage will not provide a perfect noiseless voltage.
Also, we have assumed that the actuating gate does not affect the SET.
This last assumption is realistic, however, for two reasons. First,
the resonator itself acts as a shield between the gate and the SET.
Second, since the observer knows the voltage on the feedback gate, she
can subtract that effect off the SET signal, while adding some noise.

As mentioned previously, a quantum mechanical harmonic oscillator and
a classical one are indistinguishable as long as the wave function is
Gaussian, which is the case in the present analysis.  Therefore,
although the oscillator is near the quantum mechanical ground state,
the SET measurement of position will not show any quantum behavior.
In the face of these limitations, it is a pleasant result that
experimentally obtainable situations today allow for the feedback
cooling of an resonator to the point that quantum behavior could
become distinguishable from classical behavior with an appropriate
measurement scheme.

\acknowledgments The authors would like to thank Miles Blencowe,
Alexander Korotkov, Daniel Steck, Howard Wiseman, Bernard Yerke, and
Yong Zhang for helpful conversations and suggestions. Figure 1 is
reprinted courtesy of Los Alamos Science. This research was supported
in part by the Department of Energy, under contract W-7405-ENG-36.

\appendix

\section{The Spectra of the SET shot noise and back-action}
\label{Spectra}
Here we discuss briefly how the expressions for the shot noise and
back-action of the position measurement via a SET are obtained. For
more details the reader is referred to Zhang and
Blencowe~\cite{Zhang}, (from which we obtain most of the following
expressions) and Korotkov~\cite{Korotkov94}.

The SET consists of a central island, which electrons tunnel in and
out of via junctions on either side. If one requires that the spacing
between the energy levels of the electron states on the island are
sufficiently large compared to the voltage drop across the SET, then
only two island states will be appreciably populated, these being the
states in which there are $n$ and $n+1$ electrons on the island, for
some $n$. This is because the transition rates which connect these
states to the other states are suppressed. The value of $n$ can be set
by biasing the central island. In particular, $n$ is determined by the
condition
\begin{equation}
  n < \left(\frac{C_{\ms{g}}}{e}\right) \left( V_{g} - V_{\rm ds}/2 \right) < n+1.
  \label{ncon}
\end{equation} 
As a result, we can write a master equation for the probability 
density for the occupation of the two states. Denoting this 
density by $\tilde{\sigma} = (\sigma(n),\sigma(n+1))^T$, we have
\begin{equation}
  \frac{d \tilde{\sigma}}{dt} = \left( \begin{array}{cc} -a(n) & b(n+1) \\ 
                                a(n)& -b(n+1) \end{array} \right) \tilde{\sigma} ,
                                \label{SETME}
\end{equation} 
where $a(n)$ is the transition rate from $n$ to $n+1$, and $b(n+1)$ is 
the transition rate from $n+1$ to $n$. 

If we denote the tunneling rates into the island across the source
junction and the drain junction (see Fig.~\ref{figSET}) as $a_-(n)$
and $a_+(n)$ respectively (the plus and minus subscripts record
whether the tunneling event has a positive or negative contribution to
the SET current), and out of the island as $b_+$ and $b_-$,
respectively, then
\begin{eqnarray}
  a(n) & = & a_+(n) + a_-(n) , \\
  b(n+1) & = & b_+(n+1) + b_-(n+1) .
\end{eqnarray}
It is also useful to define
\begin{eqnarray}
  f(n) & = & a_+(n) - a_-(n) , \\
  g(n+1) & = & b_+(n+1) - b_-(n+1) . 
\end{eqnarray}
In what follows we will repress the arguments of these functions, so
that $a\equiv a(n)$, $b\equiv b(n+1)$ etc. The solution to the master
equation is
\begin{eqnarray}
  \tilde{\sigma}(t) =\left[ \left( \begin{array}{rr} b & b \\ a & a \end{array} \right) 
                  +       \left( \begin{array}{rr} a & -b \\ -a & b \end{array} \right)
                  e^{-(a+b)t} \right] \frac{\tilde{\sigma}(0)}{(a+b)}. 
\end{eqnarray}
From this it is straightforward to calculate the average steady state
current flowing through the SET, the noise spectra of the current,
$S_{\mbox{\scriptsize I}}(\omega)$, along with that of an arbitrary
function, $\phi(n)$, of the island electron number $S_\phi(\omega)$,
and their mutual correlation spectrum, $C(\omega)$.  The average
current is
\begin{eqnarray}
 I = (ag + bf)/(a+b) ,
 \label{current}
\end{eqnarray}
and the spectra are 
\begin{equation}
  S_\phi(\omega) = \frac{2ab}{(a+b)} \frac{[\phi(n)-\phi(n+1)]^2}{(a+b)^2 + \omega^2}
\label{Sphi}
\end{equation}
\begin{equation}
  S_{\mbox{\scriptsize I}}(\omega) = \frac{2e^2C^2}{(a+b)C_{\mbox{\tiny $\Sigma$}}^2} 
                 \left[ ab + \frac{(f-g)(a^2g - b^2f)}{(a+b)^2 + \omega^2} \right]
  \label{SI}
\end{equation}
\begin{equation}
  C^2(\omega) = \frac{(ag + bf)^2(a-b)^2 + \omega^2(ag - bf)^2} 
                     { 4ab \left[ ab \left[ (a+b)^2 + \omega^2 \right] 
                         + (f-g)(a^2g - b^2f)\right]}
  \label{C2}
\end{equation}
Since the force from the island on the resonator is given by~\cite{Zhang}
\begin{equation}
  F = \frac{C_{\mbox{\scriptsize g}}(2C - C_{\mbox{\scriptsize g}})}{2 C_{\mbox{\scriptsize $\Sigma$}}^3 d}[C(V_{\mbox{\scriptsize ds}} - 2V_{\ms{g}}) - ne]^2 ,
\end{equation}
using Eq.~(\ref{Sphi}) we have 
\begin{equation}
  S_{\mbox{\scriptsize F}}(\omega) = \frac{ab e^2 C_{\mbox{\scriptsize g}}^2(2C - C_{\mbox{\scriptsize g}})^2}{2(a+b) C_{\mbox{\scriptsize $\Sigma$}}^6 d^2} \frac{[C (V_{\mbox{\scriptsize ds}} - 2V_{\ms{g}}) - e(2n+1)]^2}
                     {(a+b)^2 + \omega^2}
  \label{SF}
\end{equation}

Recall that in deriving these expressions we require that the two-level 
approximation is valid, and this demands that
\begin{eqnarray}
 V_{\ms{ds}} & \ll & e/C_\Sigma , \\
 k_B T & \ll & e V_{\ms{ds}} .
\label{apprates}
\end{eqnarray}
The tunneling rates are given by 
\begin{eqnarray}
    a_{\pm}(n)&=&\frac{(\Delta n\pm\tilde{V}_{\rm
    ds})/(R_{j}C_{\Sigma})}{1-\exp\left[-(\Delta n\pm\tilde{V}_{\rm
    ds})/\tilde{T}\right]}, \label{appcon1} \nn \\ 
    b_{\pm}(n+1)&=&\frac{(-\Delta n\pm\tilde{V}_{\rm
    ds})/(R_{j}C_{\Sigma})}{1-\exp\left[-(-\Delta
    n\pm\tilde{V}_{\rm ds})/\tilde{T}\right]}, \nn 
    \label{appcon2}
\end{eqnarray}
where 
\begin{eqnarray}
    \Delta n&=&\frac{C_{g}V_{g}}{e} -\frac{C_{g}V_{\rm ds}}{2e} -n - 
    \frac{1}{2},\cr
     \tilde{V}_{\rm ds}&=&\frac{C_{\Sigma}V_{\rm ds}}{2e},\cr
      \tilde{T}&=&\frac{C_{\Sigma}k_{B}T}{e^{2}}.
      \label{dimdef}
\end{eqnarray}
Note that the condition which determines $n$ (Eq.~(\ref{ncon})) is 
equivalent to $-0.5<\Delta n<0.5$.

From the expressions for the noise spectra we see that both sources of
noise are effectively white (independent of $\omega$) so long as
$\omega^2$ is much less than $[a(n)+b(n+1)]^{2}$. If this is the case
then the simple quantum theory of continuous position measurement
presented in the main body of the paper provides a good model for the
SET measurement. Note that the actual back-action noise on the
position of the resonator is the force noise filtered through the
resonator spectral function. This is therefore
\begin{equation}
    S_{X}^{F}(\omega)=\frac{S_{F}(\omega)/m_{\rm 
    eff}^{2}}{(\omega^{2}-\omega_{0}^{2})^{2}+
    \omega^{2}\omega_{0}^{2}/Q^{2}},
    \label{SXF}
\end{equation}
and has the same form as that predicted using the quantum mechanical
model (Eq.~(\ref{spectrum})), so long as the force noise is white.

We must therefore evaluate $[a(n)+b(n+1)]^{2}$ for the range of
parameters of interest, and verify that it is much larger than
$\omega^2$ over the relevant frequency range.  First we note that the
form of the spectral equations is such that they are periodic in the
gate voltage.  That is, the values of $a(n)$ and $b(n+1)$ depend only
on $\Delta n$, not on the particular value of $n$ in question.  As a
result we merely need evaluate $[a(n)+b(n+1)]^{2}$ for a single value
of $n$, and check all values of $\Delta n$ between -0.5 and 0.5.

Substituting in realistic parameter values (those that we use in our
examples in the body of paper) in Eqs.(\ref{appcon1}) and
(\ref{appcon2}), we find that, regardless of the value of $\Delta n$,
\begin{eqnarray}
[a(n)+b(n+1)] &\geq& 2 \times 10^{10} 
\end{eqnarray}
for the range of initial temperatures that we consider, and this is
much greater than the range of $\omega$ relevant for the dynamics of
the resonator, as required. Thus, we can drop $\omega$ from the
expressions for the spectra, Eqs.(\ref{SI}), (\ref{SF}) and
(\ref{C2}), and use these to determine the parameters $k$, $\beta$ and
$\kappa$ in the model of the quantum position measurement.


\begin{thebibliography}{99}

\bibitem{Experiments} 
   A.N. Cleland and M.L. Roukes, Appl. Phys. Lett. {\bf 69}, 2653 (1996).
   
\bibitem{Armour} 
   See, e.g., A.D. Armour, M.P. Blencowe, and K.C. Schwab, 
   Phys. Rev. Lett. {\bf 88}, 148301 (2002).
   
\bibitem{Korotkov94} 
   A.N. Korotkov, Phys. Rev. B {\bf 49}, 10381 (1994).
   
\bibitem{Hanke94} 
   U. Hanke {\em et al.}, Appl. Phys. Lett. {\bf 65}, 1847 (1994). 
   
\bibitem{Shnirman}
   A. Shnirman and G. Sch\"on, Phys. Rev. B {\bf 57}, 15400 (1998). 

\bibitem{Doherty99} 
   A.C. Doherty and K. Jacobs, Phys. Rev. A {\bf 60}, 2700 (1999).

\bibitem{Jacobs}
   O.L.R. Jacobs, `Introduction to Control Theory' 
                  (Oxford University Press, Oxford, 1993).

\bibitem{Zhang} 
   Y. Zhang and M.P. Blencowe, J. App. Phys. {\bf 91}, 4249 (2002),
                               Eprint: cond-mat/0109412.
                               
\bibitem{Korotkov01} A.N. Korotkov, Phys. Rev. B {\bf 63}, 115403 (2001).
                     In this article Korotkov also considers the application of 
                     feedback control to a single qubit. See also 
                     R. Ruskov and A.N.Korotkov, Phys. Rev. B {\bf 66}, 041401 (2002).
   
\bibitem{Wiseman93}
   H.M. Wiseman and G.J. Milburn, Phys. Rev. A {\bf 47}, 642 (1993).
    
\bibitem{Milburn94} 
   G.J. Milburn, K. Jacobs, and D.F. Walls, Phys. Rev. A {\bf 50}, 5256 (1994).
   
\bibitem{Caves87}
   C.M. Caves and G.J. Milburn, Phys. Rev. A {\bf 36}, 5543 (1987).
   
\bibitem{Jacobs98}
   K. Jacobs and P.L. Knight, Phys. Rev. A {\bf 57} 2301 (1998).

\bibitem{QNoise}
   C.W. Gardiner, {\em Quantum Noise} (Springer, Berlin, 2000). 
   
\bibitem{CCR}
   A.O. Caldeira, H.A. Cerdeira, and R. Ramaswamy, 
                  Phys. Rev. A {\bf 40}, 3438 (1989).

\bibitem{HuVit}
   See, e.g., B.L.~Hu, J.P.~Paz and Y.~Zhang,
              Phys. Rev. D {\bf 45}, 2843 (1992); {\bf 47}, 1576 (1993);
   V.~Giovannetti and D.~Vitali, 
              Phys. Rev. A {\bf 63}, 023812 (2001).
                    
\bibitem{Schoeller} H. Schoeller and G. Sch\"{o}n, 
                       Phys. Rev. B {\bf 50}, (1994).
                    
\bibitem{HWpriv}
   H.M. Wiseman, private communication (2002).  
                    
\bibitem{Blen_Wyb}
   M.P. Blencowe and M.N. Wybourne, App. Phys. Lett. {\bf 77}, 3845 (2000).
                                     
\bibitem{Doherty00}
   A.C. Doherty, S. Habib, K. Jacobs, H. Mabuchi and S.M. Tan, 
                 Phys. Rev. A {\bf 62} 012105 (2000). 
                 
\end{thebibliography}
\end{document}